\documentclass[conference]{IEEEtran}

\usepackage{cite}

\ifCLASSINFOpdf
\else
\fi

\usepackage{amsthm}
\usepackage{amsmath}
\usepackage{amssymb}
\usepackage{color}
\usepackage{graphicx}
\usepackage{fancyhdr}

\DeclareMathOperator*{\argmin}{arg\,min}
\DeclareMathOperator*{\diag}{diag}

\newtheorem{theorem}{Theorem}

\begin{document}

\title{With or Without Replacement? Improving Confidence in Fourier Imaging}

\author{\IEEEauthorblockN{Frederik Hoppe\IEEEauthorrefmark{1},
Claudio Mayrink Verdun\IEEEauthorrefmark{2},
Felix Krahmer\IEEEauthorrefmark{3}, 
Marion I. Menzel\IEEEauthorrefmark{4} and
Holger Rauhut\IEEEauthorrefmark{5}}
\IEEEauthorblockA{\IEEEauthorrefmark{1} Chair of Mathematics of Information Processing, RWTH Aachen University, Germany\\ 
}
\IEEEauthorblockA{\IEEEauthorrefmark{2}Harvard John A. Paulson School of Engineering and Applied Sciences, Harvard University, USA}
\IEEEauthorblockA{\IEEEauthorrefmark{3}TUM School of Computation, Information and Technology, Technical University Munich, Germany, and\\
Munich Center for Machine Learning, Germany}
\IEEEauthorblockA{\IEEEauthorrefmark{4}Faculty of Electrical Engineering and Information Technology, TH Ingolstadt, Germany,\\
GE HealthCare, Munich, Germany, and TUM School of Natural Sciences, Munich, Germany}
\IEEEauthorblockA{\IEEEauthorrefmark{5}Department of Mathematics, LMU Munich, Germany, and Munich Center for Machine Learning, Germany}}

\maketitle

\begin{abstract}
Over the last few years, debiased estimators have been proposed in order to establish rigorous confidence intervals for high-dimensional problems in machine learning and data science. The core argument is that the error of these estimators with respect to the ground truth can be expressed as a Gaussian variable plus a remainder term that vanishes as long as the dimension of the problem is sufficiently high. Thus, uncertainty quantification (UQ) can be performed exploiting the Gaussian model. Empirically, however, the remainder term cannot be neglected in many realistic situations of moderately-sized dimensions, in particular in certain structured measurement scenarios such as Magnetic Resonance Imaging (MRI). This, in turn, can downgrade the advantage of the UQ methods as compared to non-UQ approaches such as the standard LASSO.
In this paper, we present a method to improve the debiased estimator by sampling without replacement.
Our approach leverages recent results of ours on the structure of the random nature of certain sampling schemes showing how a transition between sampling with and without replacement can lead to a weighted reconstruction scheme with improved performance for the standard LASSO. In this paper, we illustrate how this reweighted sampling idea can also improve the debiased estimator and, consequently, provide a better method for UQ in Fourier imaging.

\end{abstract}

\IEEEpeerreviewmaketitle

\section{Introduction}
\def\thefootnote{}\footnotetext{\copyright 2024 IEEE. Personal use of this material is permitted. Permission from IEEE must be obtained for all other uses, in any current or future media, including reprinting/republishing this material for advertising or promotional purposes, creating new collective works, for resale or redistribution to servers or lists, or reuse of any copyrighted component of this work in other works.}\def\thefootnote{\arabic{footnote}}
High-dimensional models have become ubiquitous across various scientific disciplines, with notable prominence in fields where machine learning or signal processing techniques are used. Given their extensive application, it has become crucial to accurately assess the uncertainty surrounding the solutions to these models. This necessity arises from the inherent presence of noise in the data, which directly influences the solutions obtained by solving such models with a certain optimization strategy.

Quantifying uncertainty in high-dimensional regression models like the LASSO poses a significant challenge. These estimators often introduce a bias in order to not compromise the variance. This results in biased estimates that are unsuitable for making inferences on model coefficients, as shown by asymptotic results for fixed dimension derived in \cite{fu2000asymptotics} for the LASSO estimator. Moreover, the LASSO introduces bias by shrinking all the coefficients towards zero, which helps in variable selection and prevents overfitting in high-dimensional settings, but this shrinkage can lead to underestimation of true effect sizes and compromises the ability to draw accurate statistical inferences about individual coefficients. The problem of uncertainty quantification (UQ) for high-dimensional regression models received a lot of attention recently since, in the case of sparse regression, a few papers initiated a post-selection debiased approach to rigorously obtain confidence intervals for the LASSO coefficients. These methods have great potential to guide decision-making in critical applications like medical imaging \cite{10096320}.

The main idea is that a modification of the LASSO based on its KKT conditions, the so-called \emph{debiased LASSO}, yields a solution that approximately follows a Gaussian distribution. Thus, confidence intervals for the coefficients can be deduced. A key feature of this approach is that it exploits sparsity constraints of the underlying model. Under such assumptions, previous works rigorously quantify the performance for measurement systems that are subgaussian or given by a bounded orthonormal system \cite{vandeGeer.2014,Javanmard.2014,Javanmard.2018, journal2022}.

However, in many applications, including telecommunications and medical imaging, the underlying signal is typically not sparse in the canonical basis. Therefore, in order to use sparse regression techniques for such applications, one needs to work with a sparsifying transform, either a general-purpose representation system such as a wavelet basis or a learned dictionary. In this case, the debiased results established for UQ, e.g., \cite{journal2022}, are applicable in a somewhat restricted setting. Even for the simple case of sparsity in the Haar wavelet domain \cite{burrus1998wavelets}, most theory is based on non-uniform sampling with replacement \cite{adcock2021compressive}, which can lead to many points being sampled multiple times and, consequently, a lower number of distinct samples. 

As observed in \cite{sampta2023}, this argument can also be turned around: When a certain number of distinct samples is observed, this corresponds to a sampling-with-replacement model with a larger (virtual) number of measurements provided this transformation is reflected by a reweighting in the LASSO reconstruction.
In \cite{sampta2023}, we explored the effect of this transformation on the reconstruction accuracy for the standard LASSO. In this work, we demonstrate that it can also improve the UQ performance.
This is important for Fourier imaging with Haar wavelet sparsity, as too few samples can make the UQ procedure for some coefficients meaningless. This situation is even more challenging if such UQ methods are employed for learning-based methods \cite{MLSP2023}. 

\textbf{Our contribution:} This work aims to show that the reweighted scheme of \cite{sampta2023} can overcome the aforementioned problem and improve uncertainty quantification techniques for sparse estimators when the underlying ground truth is sparse in a non-trivial domain instead of the canonical basis. In particular, we show that by using a reweighted sampling without replacement scheme, we can obtain sharper debiased estimators with better convergence properties. This allows for constructing more precise confidence intervals in cases where the ground truth is sparse only after a change of basis.

\section{The Debiased LASSO}\label{sec:debiased_LASSO}

We consider measurements given by a linear model $y = A x^0 + \varepsilon$ with $s$-sparse ground truth $x^0\in\mathbb{C}^N$, measurements matrix $A\in\mathbb{C}^{m\times N}$ and complex Gaussian noise $\varepsilon\sim\mathcal{CN}(0, \sigma^2 I_{N\times N})$. The LASSO estimator $\hat{x}$ \cite{tibshirani1996regression} retrieves the signal by solving the problem
\begin{equation*}
    \argmin\limits_{x\in\mathbb{C}^N} \frac{1}{2m}\Vert Ax -y\Vert_2^2 + \lambda \Vert x\Vert_1
\end{equation*}
with regularization parameter $\lambda>0$. We assume that the matrix $A$ is normalized such that the sample covariance $\hat{\Sigma}_A:=\frac{1}{m}A^*A$ has diagonal entries of order one. Thanks to the $\ell_1$-regularization, which introduces a shrinkage of the coefficient magnitudes, the LASSO is biased \cite{giraud2021introduction}. A few works \cite{Zhang.2014, vandeGeer.2014, Javanmard.2014} established a correction to remove this bias from the LASSO, i.e.
\begin{equation*}
    \hat{x}^u = \hat{x} + \frac{1}{m}A^*(y-A\hat{x}).
\end{equation*}
The corrected estimator is called \emph{debiased LASSO}. The main achievement of the debiased estimation theory is the decomposition
\begin{equation}\label{eq:decomposition}
    \hat{x}^u - x^0 = \underbrace{A^*\varepsilon/m}_{=:W} + \underbrace{(\hat{\Sigma}_A - I_{N\times N})(x^0-\hat{x})}_{=:R},
\end{equation}
with a Gaussian term $W \sim \mathcal{CN}(0,\sigma^2\hat{\Sigma}_A)$ and a remainder term $R$, that vanishes asymptotically, i.e. when $m\to\infty$ and $N=N(m)\to\infty$ such that $N/m$ is constant and $\frac{s_0\log^2 N}{m}\to 0$; see \cite[Remark 1.2]{Javanmard.2018}. Thus, the debiased LASSO is asymptotically Gaussian with mean $x^0$. This allows for constructing confidence intervals based on the distribution of $W$. The confidence region with significance level $\alpha\in(0,1)$ for the complex pixel value $x_i^0$ is given by
\begin{equation*}
    J_i(\alpha) = \{ z \in\mathbb{C}: \vert \hat{x}^u_i - z \vert \leq \delta_i(\alpha)\}
\end{equation*}
with radius $\delta_i(\alpha) = \frac{\hat{\sigma}(\hat{\Sigma}_A)_{ii}^{1/2}}{m}\sqrt{\log(1/\alpha)}$. A detailed derivation of the confidence regions can be found, e.g., in \cite{journal2022}.

\section{Sampling schemes}

The theory of compressive sampling for image retrieval requires that the measurement operator is well-behaved on certain sets, e.g., on a union of subspaces. Such a notion is mathematically described by concepts such as incoherence, the restricted isometry property, or the nullspace property \cite{Foucart.2013}. In the case when the measurement matrix is given by a subsampled Fourier matrix $F_{\Omega}$, which is the measurement scheme employed in MRI, it is known that it has the restricted isometry property (RIP) with high probability provided that its rows are sampled uniformly at random \cite[Thm. 1.1 and 2.3]{brugiapaglia2021sparse}.

However, in cases when a sparsifying transform such as the Haar wavelet is incorporated and hence, not the signal $x^0\in\mathbb{C}^N$, but $z^0 = Hx^0$ is $s$-sparse, the new measurement operator $A=F_{\Omega}H^*$ is coherent, see \cite[Chapter 11]{adcock2021compressive}. In this case, a non-uniform sampling strategy must be employed to guarantee that the measurement operator is well-behaved. The following result from \cite{krahmer2013stable} shows that non-uniform sampling ensures that the Fourier-Wavelet measurement scheme fulfills the RIP. We refer to \cite[Definition 6.1.]{Foucart.2013} for the definition and a discussion of the RIP.

\begin{theorem}\cite[Section V, Theorem 1]{krahmer2013stable}\label{thm:wBOS}
Let $\Phi=\{\varphi_j\}_{j=1}^N$ and $\Psi =\{\psi_k\}_{k=1}^N$ be orthonormal bases of $\mathbb{C}^N$. Assume the local coherence of $\Phi$ with respect to $\Psi$  is pointwise bounded by the function $\kappa$, that is  $ \sup\limits_{1\leq k\leq N} |\langle \varphi_j, \psi_k\rangle| \leq \kappa_j$. 
Let $s\gtrsim \log(N)$, suppose
\begin{equation*}
m \gtrsim \delta^{-2} \|\kappa \|_2^2 s \log^3(s) \log(N),
\end{equation*}
and choose $m$ (possibly not distinct) indices $j \in \Omega \subset [N]$ i.i.d. from the probability measure $\nu$ on $[N]$ given by
\begin{equation}\label{eq:non_uniform_prob_measure}
\nu(j) = \frac{\kappa^2_j}{\|\kappa \|_2^2 }.
\end{equation}
Consider the matrix $A \in \mathbb{C}^{m \times N}$ with entries
\begin{equation*}
A_{j,k} = \langle \varphi_j, \psi_k\rangle, \quad j \in \Omega, k \in [N],
\end{equation*}
and consider the diagonal matrix $D = \operatorname{diag}(d) \in \mathbb{C}^{m}$ with $d_{j} = \| \kappa \|_2 / \kappa_j$.
Then, with a probability of at least 
$1-N^{-c \log^3(s)},$ the restricted isometry constant $\delta_s$ of the preconditioned matrix $\frac{1}{\sqrt{m}} D A$ satisfies $\delta_s \leq \delta$.
\end{theorem}
The rows of $F$ are now sampled with replacement according to the non-uniform probability measure \eqref{eq:non_uniform_prob_measure} and the measurement matrix $F_{\Omega}H^*$ is normalized through the preconditioning diagonal matrix $D$, that depends on the measure $\nu$.

The debiased LASSO applied to this problem with measurement matrix $B:=DF_{\Omega}H^*$ yields a decomposition in the sense of \eqref{eq:decomposition}
\begin{equation*}
    \hat{z} - z^0 = \underbrace{ (D^2F_{\Omega}H^*)^*\varepsilon/m }_{=:W^z} + \underbrace{ (\hat{\Sigma}_B - I_{N\times N})(z^0 - \hat{z}) }_{=:R^z},
\end{equation*}
where $\hat{z}$ denotes the LASSO for the equivalent model $Dy = DF_{\Omega}H^*z^0 + D\varepsilon$.

In practice, however, this gives rise to a \textbf{tradeoff}: If we sample according to measure $\eqref{eq:non_uniform_prob_measure}$ with replacement, then many rows will be sampled more than once with high probability. If we sample without replacement, in contrast, which seems much more natural from the perspective of maximizing acquired information, Theorem \ref{thm:wBOS}
does not apply.

When considering the debiased LASSO, sampling from $\nu$ without replacement has another disadvantage: the matrix $\hat{\Sigma}_B$ of $R^z$ has diagonal entries of different sizes, which makes uniform normalization impossible and hence slows down the asymptotic convergence of $R^z$.

We overcome this problem by considering \emph{reweighted sampling without replacement} \cite{sampta2023}, which can be interpreted as transforming the distinct samples into a virtual model of sampling with replacement. Computationally, one independently draws samples $\omega_1,\hdots\,\omega_{n}$ with replacement from a probability measure until obtaining $m$ distinct samples, which one physically acquires. The counts $\gamma_1\hdots,\gamma_m$, how often the samples occur in the virtual model, are recorded for the reconstruction procedure, which can be taken into account to mimic a model with replacement with $n=\sum_{i=1}^m\gamma_i$ samples.

\section{Improving the Debiased LASSO's Confidence}
We can now leverage the sampling strategy to construct an unbiased LASSO with better recovery and inference properties than the standard construction. This standard approach is a direct application of the debiasing step for the LASSO as described in Section \ref{sec:debiased_LASSO}. Our new approach tailors the debiasing step to a Haar-transformed signal using reweighted sampling without replacement. This bridges sampling without replacement (used, e.g., in practical MRI scenarios) with theoretical recovery guarantees for sampling with replacement that are given, e.g., in Theorem \ref{thm:wBOS}.

\subsection{Standard Debiasing}
We select the rows indexed by the set $\Omega\in\mathbb{N}^m$ with or without replacement and obtain a subsampled Fourier matrix $F_{\Omega}$. After solving the LASSO
\begin{equation*}
    \argmin\limits_{z\in\mathbb{C}^p} \frac{1}{2m}\Vert y - F_{\Omega}H^* z^0\Vert_2^2 + \lambda \Vert z\Vert,
\end{equation*}
we construct the debiased LASSO by adding
\begin{equation*}
    \hat{z}^u = \hat{z} + \frac{1}{m} (F_{\Omega}H^*)^*(y - F_{\Omega}H^*\hat{z}).
\end{equation*}
This gives us, in the Haar domain, the decomposition
\begin{align*}
    \hat{z}^u - z^0 = \underbrace{(F_{\Omega}H^*)^*\varepsilon/m}_{=:W^z} + \underbrace{(H\hat{\Sigma}_FH^* - I_{N\times N})(z^0-\hat{z})}_{=:R^z}
\end{align*}
with $\hat{\Sigma}_F = \frac{1}{m}F_{\Omega}^*F_{\Omega}$. In the image domain, we obtain
\begin{align*}
     \hat{x}^u-x^0 = \underbrace{F_{\Omega}^*\varepsilon/m}_{=:W^x} + \underbrace{(\hat{\Sigma}_F - I_{N\times N})(x^0-H^*\hat{z})}_{=:R^x}.
\end{align*}

\subsection{Reweighting Sampling Without Replacement Debiasing}
Our more sophisticated approach takes into account that sampling without replacement but with reweighting yields better numerical performance, as described, e.g., by numerical experiments in \cite{sampta2023}. Following \cite{sampta2023}, we define a reweighted version of the LASSO and explain why debiasing this LASSO estimator overcomes the tradeoff mentioned above. Although we restrict ourselves to the Haar domain, the result is transferrable into the image domain by exploiting the fact that the Haar transform is an isometry with respect to the $\ell_2$-norm.

\begin{theorem}\label{thm:our_method}
    Assume the setting of Theorem \ref{thm:wBOS}. Let $\gamma_1,\hdots, \gamma_m$ be the count records of the reweighted sampling without replacement, $C=\diag(\sqrt{\gamma_1},\hdots, \sqrt{\gamma_m})$ and $D\in C^{m\times m}$ as defined in Theorem \ref{thm:wBOS}.
    Let $\Omega\in\mathbb{N}^m$ be drawn from $\nu$ without replacement and let $n = \sum_{i=1}^m\gamma_i$. Denote by $\hat{z}$ the LASSO solution of
    \begin{equation}\label{eq:reweighted_deb_lasso}
    \min\limits_{z\in\mathbb{C}^N}\frac{1}{2n}\Vert  CD (F_{\Omega}H^*x^0-y)\Vert_2^2+\lambda\Vert z\Vert_1,
    \end{equation}
    and by $\tilde{\hat{z}}$ the one of
    \begin{equation}\label{eq:standard_deb_lasso}
    \min\limits_{z\in\mathbb{C}^N}\frac{1}{2n}\Vert  \tilde{D} (F_{\tilde{\Omega}}H^*x^0-y)\Vert_2^2+\lambda\Vert z\Vert_1,
    \end{equation}
    where $\tilde{\Omega}\in\mathbb{N}^n$ is sampled with replacement, and $\tilde{D}\in\mathbb{C}^{n\times n}$ the corresponding diagonal matrix. Then, it holds that
    \begin{equation*}
        (CDF_{\Omega}H^*)^*CDF_{\Omega}H^* = (\tilde{D}F_{\tilde{\Omega}}H^*)^*\tilde{D}F_{\tilde{\Omega}}H^*.
    \end{equation*}
    This means that the remainder term $R^z$ of the debiased LASSO $\hat{z}^u$ derived from \eqref{eq:reweighted_deb_lasso}, i.e.
    \begin{equation*}
    R^z=\left((CDF_{\Omega}H^*)^*CDF_{\Omega}H^*/n - I_{N\times N}\right)(z^0-\hat{z})
    \end{equation*}
    can be interpreted as the remainder term of the debiased LASSO $\tilde{\hat{z}}^u = \tilde{\hat{z}} + \frac{1}{n}(\tilde{D}F_{\tilde{\Omega}}H^*)^*(\tilde{D}y - \tilde{D}F_{\tilde{\Omega}}H^*\tilde{\hat{z}})$, derived from \eqref{eq:standard_deb_lasso}. In particular, $\mathbb{E}[(CDF_{\Omega}H^*)^*CDF_{\Omega}H^*/n] = I_{N\times N}$. 
\end{theorem}
This theorem suggests our reweighted debiasing with $m$ distinct samples to behave like debiasing based on $\sum_{i=1}^m\gamma_i$ samples drawn with replacement. Since the sampling with replacement is only \emph{virtually} performed, it overcomes the drawback of resource-intensive sampling. With this result, we have shown that our approach takes advantage of both sampling with replacement and sampling without replacement. On the one hand, from the equivalence to sampling with replacement, we have no normalization obstacle as we had in the sampling without replacement case, and the RIP holds for the measurement matrix. On the other hand, we save resources by only subsampling $m$ distinct rows of $F$. This is of high interest, especially in MRI. 
\begin{proof}
    The model $y = F_{\Omega}H^*z^0 + \varepsilon$ is equivalent to
    \begin{equation*}\label{eq:formulation_reweighting}
        \frac{1}{\sqrt{n}} CDy = \frac{1}{\sqrt{n}} CDF_{\Omega}H^*z^0 + \frac{1}{\sqrt{n}} CD\varepsilon,
    \end{equation*}
    in the sense that the multiplication with $CD/\sqrt{n}$ is bijective. From this, we derive the debiased LASSO for $z^0$ as
\begin{equation*}
    \hat{z}^u = \hat{z} + \frac{1}{n} (CDF_{\Omega}H^*)^*(CDy - CDF_{\Omega}H^*\hat{z}).
\end{equation*}
and the decomposition as
\begin{align*}
    &\hat{z}^u - z^0 = \underbrace{ \frac{1}{n}(CDF_{\Omega}H^*)^*CD\varepsilon }_{=:W^z}\\
    &+ \underbrace{ \left(\frac{1}{n}(CDF_{\Omega}H^*)^*CDF_{\Omega}H^* - I_{N\times N}\right)(z^0-\hat{z}) }_{=:R^z}.
\end{align*}
Now, it holds that
\begin{equation*}
    (CDF_{\Omega}H^*)^*CDF_{\Omega}H^* = H\left(\sum\limits_{i=1}^m d_i^2\cdot c_i^2 f_{\omega_i} f_{\omega_i}^*\right)H^*,
\end{equation*}
where $f_{\omega_i}$ denotes the $\omega_i$-th row. Since $c_i^2=\gamma_i$ is the number of counts for the $\omega_i$-th row it can be written as
\begin{align*}
    &\sum\limits_{i=1}^m \underbrace{(d_i^2f_{\omega_i} f_{\omega_i}^*+\hdots+d_i^2f_{\omega_i} f_{\omega_i}^*)}_{\gamma_i-\text{times}}
    =\sum\limits_{j=1}^n \tilde{d}_j^2\cdot f_{\tilde{\omega}_j} f_{\tilde{\omega}_j}^* \\
    &= (\tilde{D} F_{\tilde{\Omega}})^*(\tilde{D}F_{\tilde{\Omega}})
\end{align*}
with $f_{\tilde{\omega}_1}=\hdots = f_{\tilde{\omega}_{\gamma_1}}=f_{\omega_1}$, $\tilde{d}_1=\hdots =\tilde{d}_{\gamma_1}=d_1$ $,\hdots,$ $f_{\tilde{\omega}_{n-\gamma_m+1}}=\hdots =f_{\tilde{\omega}_n}= f_{\omega_m}$, $\tilde{d}_{n-\gamma_m+1}=\hdots=\tilde{d}_{n}=d_m$. This is the same as having $n$ measurements sampled with replacement when deriving the debiased LASSO \eqref{eq:standard_deb_lasso} from the model $\tilde{D}y = \tilde{D}F_{\tilde{\Omega}}H^*z^0 + \tilde{D}\varepsilon$ where $\tilde{\Omega}$ contains the indices $\omega_i$ with multiplicity $\gamma_i$, i.e. $\tilde{\omega_1},\hdots,\tilde{\omega_n}$. In particular, we obtain
\begin{align*}
    &\mathbb{E}\left[\frac{(CDF_{\Omega}H^*)^*CDF_{\Omega}H^*}{n} \right]= H \mathbb{E}\left[\frac{1}{n}\sum\limits_{i=1}^m d_i^2 c_i^2 f_{\omega_i} f_{\omega_i}^*\right]H^*\\
    &= H\mathbb{E}\left[\frac{1}{n}\sum\limits_{j=1}^n \tilde{d}_j^2 f_{\tilde{\omega}_j} f_{\tilde{\omega}_j}^*\right]H^*
    =\mathbb{E}\left[\frac{(\tilde{D}F_{\tilde{\Omega}}H^*)^*(\tilde{D}F_{\tilde{\Omega}}H^*)}{n}\right]\\
    & = I_{N\times N},
\end{align*}
where the last equality holds since $\frac{1}{\sqrt{n}}\tilde{D}F_{\tilde{\Omega}}H^*$ is a random sampling matrix associated to a BOS as shown in \cite{krahmer2013stable}.
\end{proof}

\section{Numerical Experiments}
\begin{figure}
    \centering
    \includegraphics[width = 0.3\columnwidth]{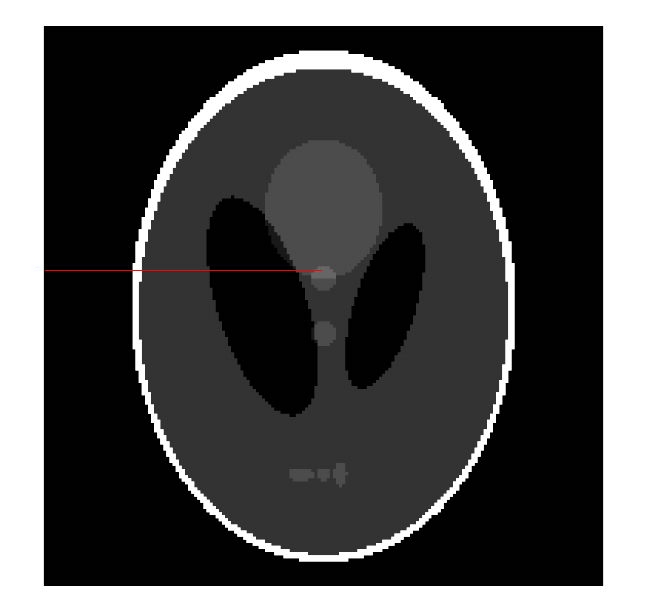}
    \caption{Modified Shepp-Logan phantom. The marked red line shows the pixel for which Figure \ref{fig:conf_int} displays the confidence intervals.}
    \label{fig:marked_line}
\end{figure}

In this section, we compare the standard debiasing without replacement against our method, \emph{the reweighting sampling without replacement debiasing}. First, we do not use standard debiasing with replacement due to the large number of required samples. Second, the standard debiasing without replacement strategy suffers from a missing uniform normalization of $\hat{\Sigma}_B$. Theorem \ref{thm:our_method}
shows that our method overcomes this issue while having the same sample complexity, in terms of the required $m$, as the sampling without replacement method. In the experiments, we simulate the MRI process. For the reconstruction, we use the solver TFOCS \cite{becker2011templates}, which is a first-order solver for a convex conic problem (the chosen algorithm was Auslender and Teboulle's single-projection method \cite{auslender2006interior}). As a ground truth, we use a modified version of the Shepp-Logan Phantom (see Figure \ref{fig:marked_line}) denoted in a vectorized version by $x^0\in\mathbb{C}^N$ with $N=32768$. The underlying MRI model reads as $y = F_{\Omega}x^0 + \varepsilon$, where $\varepsilon\in\mathbb{C}^m$ is complex Gaussian noise with covariance structure $\sigma^2 I_{m\times m}$. The index set $\Omega\in\mathbb{N}^m$ is sampled from the probability measure in \eqref{eq:non_uniform_prob_measure} without replacement for the standard debiased LASSO and with reweighting without replacement for our method. Then, $x^0$ is Haar transformed to $z^0 = Hx^0$. Both debiasing approaches are performed for $\lambda = k\cdot \lambda_0$ with $k \in\{5,10,15,20,25\}$ and $\lambda_0:= \frac{\sigma}{m}(2+\sqrt{12\log(N))}$. The noise level in the standard approach is chosen, such that the signal-to-noise ratio is $\frac{\Vert \varepsilon\Vert_2}{\Vert F_{\Omega}H^*z^0\Vert_2}\approx 0.045$. For comparison reasons, in the reweighting scenario, it is also chosen as $\frac{\Vert CD \varepsilon \Vert_2}{\Vert CD F_{\Omega}H^*z^0\Vert_2}\approx 0.045$. In practice, the noise level can be precisely measured \cite{AjaFernandez.2016}. Therefore, the assumption of known $\sigma$ does not limit our experiments and allows us to focus on the comparison between the two methods.

\begin{figure}
    \centering
    \includegraphics[width = .5\columnwidth]{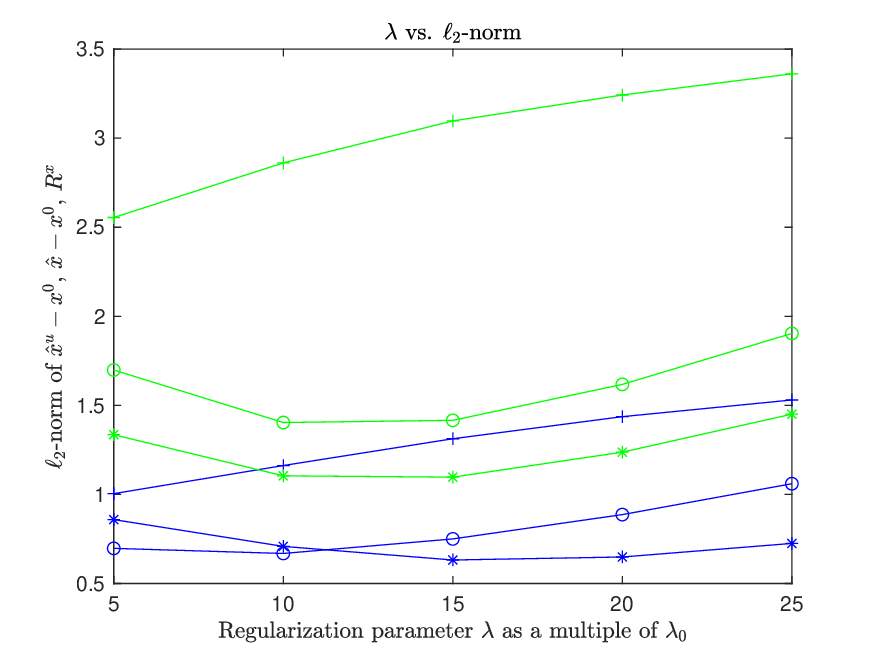}
    \caption{Quantitative comparison between the two methods: The blue markers are the values for the reweighting approach, and the green for the straightforward approach. The x-axis represents the dependency on $\lambda$ as a multiply of $\lambda_0$. The y-axis shows the $\ell_2$-norm of the following quantities: $\hat{x}-x^0$ (circle), $\hat{x}^u-x^0$ (plus) and $R^x$ (star). Note that the same values apply for $z$ since the Haar transform is an isometry w.r.t. the $\ell_2$-norm.}
    \label{fig:diagram}
\end{figure}

\begin{figure}
    \centering
    \includegraphics[width = 0.6\columnwidth]{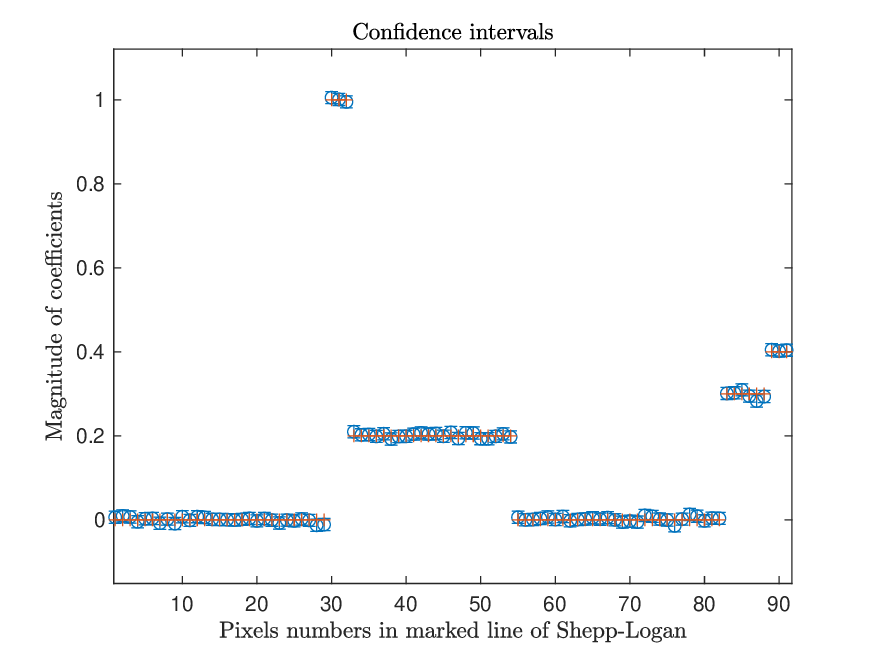}
    \caption{Confidence intervals (blue errorbars) with debiased LASSO (blue circles) for underlying image pixels (red pluses) along the marked red line in Figure \ref{fig:marked_line} shown for one realization of the subsampling and noise. Here, we chose $\lambda = 15 \lambda_0$.}
    \label{fig:conf_int}
\end{figure}

We compute the average of the estimator errors as well as the remainder and Gaussian term and show the results in Table \ref{tab:shepp_logan_straightforward} and \ref{tab:shepp_logan_reweighting} for the standard and reweighting debiased LASSO, respectively. Due to the isometry property, the $\ell_2$-norm of the quantities are the same in the Haar and image domain. The $\ell_{\infty}$-norm is considered since we aim for \emph{pixelwise} confidence interval. The error of the LASSO, the debiased LASSO, and the remainder term is significantly smaller in the reweighting setting than in the standard setting. Their dependency on $\lambda$ is displayed in Figure \ref{fig:diagram}. In addition, the Gaussian term, which is independent of $\lambda$, is much smaller in the reweighting scenario, leading to sharper confidence intervals. Here, to achieve a small ratio $\frac{\Vert R\Vert_2}{\Vert W\Vert_2}$, and hence a dominating Gaussian term $W$, a suitable choice is, e.g., $\lambda = 15 \lambda_0$. The resulting confidence intervals for one realization of the sampling pattern and the noise are presented in Figure \ref{fig:conf_int} for the red line in the Shepp-Logan phantom. Overall pixels, the confidence intervals contain $97.85\%$, and on the support, they contain $97.77\%$. 

\vspace{-.3cm}
\begin{table}[tb]
\caption{Sampling without replacement for $m=0.6p$ of Shepp-Logan phantom: Results are averaged over $l=25$ realizations of the noise $\approx 4.5\%$.}
\label{tab:shepp_logan_straightforward}
\centering
\begin{tabular}{ |l|c|c|c|c|c|c|c } 
 \hline
 $\lambda$ & 5$\lambda_0$ & 10$\lambda_0$ & 15$\lambda_0$ & 20$\lambda_0$ &25$\lambda_0$ \\
 \hline
 $\Vert \hat{z}-z^0\Vert_{\infty}$ & 0.0675 & 0.0718 & 0.0767 & 0.0885 & 0.0976\\
 \hline
 $\Vert \hat{x}-x^0\Vert_{\infty}$  & 0.0638 & 0.0697 & 0.0807 & 0.0953 & 0.1107\\
 \hline
 $\Vert \hat{x}-x^0\Vert_2$  & 1.6986 & 1.4038 & 1.460 & 1.6178 & 1.9039 \\
 \hline
 $\Vert \hat{z}^u - z^0\Vert_{\infty}$ & 0.0618 & 0.0621 & 0.0607 & 0.0611 & 0.0648\\
 \hline
 $\Vert \hat{x}^u - x^0\Vert_{\infty}$ & 0.0597 & 0.0591 & 0.0616 & 0.0620 & 0.0637\\
 \hline
 $\Vert \hat{x}^u - x^0\Vert_2$  & 2.5546 & 2.8613 & 3.0969 & 3.2424 & 3.3611\\
 \hline
 $\Vert R^z\Vert_{\infty}$  & 0.0344 & 0.0314 & 0.0322 & 0.0365 & 0.0407\\
 \hline
$\Vert R^x\Vert_{\infty}$  & 0.0308 & 0.0275 & 0.0303 & 0.0342 & 0.0403\\
 \hline
 $\Vert R\Vert_2$ & 1.3353 & 1.1044 & 1.0970 & 1.2374 & 1.4507 \\
 \hline
 $\Vert W^z\Vert_{\infty}$ & 0.0667 & 0.0674 & 0.0670 & 0.0664 & 0.0675\\
 \hline
 $\Vert W^x\Vert_{\infty}$ & 0.0595 & 0.0591 & 0.0595 & 0.0593 & 0.0588\\
 \hline
 $\Vert W\Vert_2$ & 3.2279 & 3.2286 & 3.2288 & 3.2232 & 3.2278 \\
 \hline
\end{tabular}
\end{table}

\begin{table}[tb]
\caption{Reweighted sampling without replacement for $m=0.6p$ of Shepp-Logan phantom: Results are averaged over $l=25$ realizations of the noise $\approx 4.5\%$.}
\label{tab:shepp_logan_reweighting}
\centering
\begin{tabular}{ |l|c|c|c|c|c|c|c } 
 \hline
 $\lambda$ & 5$\lambda_0$ & 10$\lambda_0$ & 15$\lambda_0$ & 20$\lambda_0$ &25$\lambda_0$ \\
 \hline
 $\Vert \hat{z}-z^0\Vert_{\infty}$ & 0.0274 & 0.0301 & 0.0329 & 0.0366 & 0.0408\\
 \hline
 $\Vert \hat{x}-x^0\Vert_{\infty}$  & 0.0268 & 0.0321 & 0.0377 & 0.0444 & 0.0509\\
 \hline
 $\Vert \hat{x}-x^0\Vert_2$  & 0.6973 & 0.6686 & 0.7503 & 0.8868 & 1.0602\\
 \hline
 $\Vert \hat{z}^u - z^0\Vert_{\infty}$ & 0.0256 & 0.0262 & 0.0273 & 0.0287 & 0.0304\\
 \hline
 $\Vert \hat{x}^u - x^0\Vert_{\infty}$ & 0.0237 & 0.0246 & 0.0258 & 0.0270 & 0.0282\\
 \hline
 $\Vert \hat{x}^u - x^0\Vert_2$  & 1.0044 & 1.1622 & 1.3127 & 1.4366 & 1.5303 \\
 \hline
 $\Vert R^z\Vert_{\infty}$  & 0.0173 & 0.0143 & 0.0129 & 0.0149 & 0.0202\\
 \hline
 $\Vert R^x\Vert_{\infty}$  & 0.0157 & 0.0129 & 0.0117 & 0.0120 & 0.0134\\
 \hline
 $\Vert R\Vert_2$ & 0.8588 & 0.7084 & 0.6319 & 0.6488 & 0.7250 \\
 \hline
 $\Vert W^z\Vert_{\infty}$ & 0.0298 & 0.0291 & 0.0297 & 0.0296 & 0.02296\\
 \hline
 $\Vert W^x\Vert_{\infty}$ & 0.0274 & 0.0274 & 0.0276 & 0.0278 & 0.0277\\
 \hline
 $\Vert W\Vert_2$ & 1.5137 & 1.5150 & 1.5112 & 1.5144 & 1.5098 \\
 \hline
\end{tabular}
\end{table}

\section{Conclusion}
This work bridges ideas from the sampling with replacement and the sampling without replacement techniques in high-dimensional. In particular, we adapted the debiased LASSO for the case when the underlying signal is sparse on a different basis. Our approach significantly decreases the estimator's error rates as compared to previous methods. In addition, our method provides sharper confidence regions, allowing for sharper uncertainty quantification.

\section*{Acknowledgment}
The authors would like to thank the German Federal Ministry of Education and Research for support through the grant  "SparseMRI3D+ (FZK 05M20WOA)".

\bibliographystyle{IEEEbib}
\bibliography{refs}

\end{document}